\documentclass[twocolumn]{aastex701}
\usepackage{natbib}
\usepackage{multirow}
\usepackage{array}
\usepackage{amsmath} 

\graphicspath{{./}{}}

\definecolor{notes}{HTML}{C70039}
\definecolor{softgreen}{HTML}{468465}
\definecolor{edits}{HTML}{BB8F00}

\newcommand{\heii}{\hbox{He\,{\sc ii}}}     
\newcommand{\oiiisemi}{\hbox{\sc O\,iii]}}  
\newcommand{\ciii}{\hbox{\sc C\,iii]}}      
\newcommand{\oii}{\hbox{\sc [O\,ii]}}     
\newcommand{\hb}{\hbox{\sc H$\beta$}}       
\newcommand{\oiii}{\hbox{\sc [O\,iii]}}     
\newcommand{\oi}{\hbox{\sc [O\,i]}}     
\newcommand{\ha}{\hbox{\sc H$\alpha$}}      
\newcommand{\nii}{\hbox{[N\,{\sc ii}]}}     
\newcommand{\sii}{\hbox{[S\,{\sc ii}]}}     
\newcommand{\siii}{\hbox{[S\,{\sc iii}]}}   
\newcommand{\pab}{\hbox{Pa$\beta$}}      

\newcommand{\feii}{\hbox{[Fe\,{\sc ii}]}} 
\newcommand{\neiii}{\hbox{[Ne\,{\sc iii]}}}  

\newcommand{\lam}{$\lambda$}
\newcommand{\unit}[1]{\ensuremath{\mathrm{\,#1}}\xspace}

\newcommand{\e}{\unit{e^{-}}}

\mathchardef\mhyphen="2D

\newlength{\dhatheight}

\usepackage{fontspec}

\shorttitle{LEGGOS: Lack of Shocks at $2<z<4$}

\begin{document}
\title{LEGGOS: A Shocking Lack of Evidence for Shocks at sub-kiloparsec Scales at $2<z<4$}

\author[0000-0001-7151-009X]{Nikko J.\ Cleri}
\altaffiliation{These authors contributed equally to this work}
\affiliation{Department of Astronomy and Astrophysics, The Pennsylvania State University, University Park, PA 16802, USA}
\affiliation{Institute for Computational \& Data Sciences, The Pennsylvania State University, University Park, PA 16802, USA}
\affiliation{Institute for Gravitation and the Cosmos, The Pennsylvania State University, University Park, PA 16802, USA}
\email[show]{cleri@psu.edu}

\author[0000-0001-6251-4988]{Taylor A. Hutchison}
\altaffiliation{These authors contributed equally to this work}
\affiliation{Astrophysics Science Division, Code 660, NASA Goddard Space Flight Center, 8800 Greenbelt Rd., Greenbelt, MD 20771, USA}
\affiliation{Department of Astronomy, University of Maryland, Baltimore County, MD 21250, USA}
\affiliation{Center for Research and Exploration in Space Science and Technology, NASA/GSFC, Greenbelt, MD 20771 USA}
\email[show]{astro.hutchison@gmail.com}

\author[0000-0003-1815-0114]{Brian Welch}
\altaffiliation{These authors contributed equally to this work}
\affiliation{International Space Science Institute, Hallerstrasse 6, 3012 Bern, Switzerland}
\email[show]{brian.welch@issibern.ch}

\author[0000-0002-3475-7648]{Gourav Khullar} 
\altaffiliation{Baum Postdoctoral Fellow for Innovative Astronomy}
\affiliation{Department of Astronomy \& the DiRAC Institute, University of Washington, Physics-Astronomy Building, Box 351580, Seattle, WA 98195-1700, USA}
\affiliation{eScience Institute, University of Washington, Physics-Astronomy Building, Box 351580, Seattle, WA 98195-1700, USA}
\affiliation{Department of Physics and Astronomy and PITT PACC, University of Pittsburgh, Pittsburgh, PA 15260, USA}
\email{cleri@psu.edu}

\author[0000-0003-1074-4807]{Matthew B. Bayliss}
\affiliation{Department of Physics, University of Cincinnati, Cincinnati, OH 45221, USA}
\email{cleri@psu.edu}

\author[0000-0003-2200-5606]{H{\aa}kon Dahle}
\affiliation{Institute of Theoretical Astrophysics, University of Oslo, P.O. Box 1029, Blindern, NO-0315 Oslo, Norway}
\email{cleri@psu.edu}

\author[0000-0001-5097-6755]{Michael Florian}
\affiliation{Steward Observatory, University of Arizona, 933 North Cherry Avenue, Tucson, AZ 85721, USA}
\affiliation{Eureka Scientific, 2452 Delmer Street Suite 100 Oakland, CA 94602-3017}
\email{cleri@psu.edu}

\author[0000-0003-1370-5010]{Michael D. Gladders}
\affiliation{Department of Astronomy and Astrophysics, University of Chicago, 5640 South Ellis Avenue, Chicago, IL 60637, USA}
\affiliation{Kavli Institute for Cosmological Physics, University of Chicago, 5640 South Ellis Avenue, Chicago, IL 60637, USA}
\email{cleri@psu.edu}

\author[0009-0001-2048-9451]{Connor Luettgenau}
\affiliation{Department of Astronomy and Astrophysics, The Pennsylvania State University, University Park, PA 16802, USA}
\email{cjluettgenau@gmail.com}

\author[0009-0000-7075-5554]{Rion Oh} 
\affiliation{Department of Astronomy \& the DiRAC Institute, University of Washington, Physics-Astronomy Building, Box 351580, Seattle, WA 98195-1700, USA}
\affiliation{Department of Physics, KAIST, Daejeon 34141, Republic of Korea}
\email{oro020@kaist.ac.kr}

\author[0000-0002-7627-6551]{Jane R. Rigby}
\affiliation{Astrophysics Science Division, Code 660, NASA Goddard Space Flight Center, 8800 Greenbelt Rd., Greenbelt, MD 20771, USA}
\email{cleri@psu.edu}

\author[0000-0002-9204-3256]{T. Emil Rivera-Thorsen}
\affiliation{The Oskar Klein Centre, Department of Astronomy, Stockholm University, AlbaNova 10691, Stockholm, Sweden}
\email{cleri@psu.edu}

\author[0000-0002-5293-3975]{Julissa Sarmiento}
\affiliation{Department of Physics and Astronomy, University of Pittsburgh, Pittsburgh, PA 15260, USA}
\email{cleri@psu.edu}

\author[0000-0002-7559-0864]{Keren Sharon}
\affiliation{Department of Astronomy, University of Michigan, 1085 S. University Ave, Ann Arbor, MI 48109, USA}
\email{cleri@psu.edu}

\collaboration{all}{the LEGGOS collaboration}

\begin{abstract}
Here we present the first systematic search for shocks in six gravitationally lensed galaxies at $2.37<z<3.625$ with JWST/NIRSpec integral field spectroscopy from the LEnsing and Galaxy Growth: Observing Substructures (LEGGOS) survey. We employ diagnostics that utilize the fluxes and kinematics of shock-sensitive rest-frame optical emission lines \hb, $\oiii~\lambda5008$, $\oi~\lambda6302$, \ha, $\nii~\lambda6585$, and $\sii~\lambda\lambda6718,6733$. We find that, on pixel, clump, and galaxy-integrated scales, the LEGGOS spectra show minimal if any evidence for shocks. The image plane pixels are $<8\%$ within the shock regions of the rest-frame optical line ratio diagnostics for any individual galaxy, and the shock-identified pixels do not show a coherent spatial structure. We also leverage MAPPINGS V shock models to infer shock velocities from the observed emission lines, and find that the distributions of inferred shock velocities are inconsistent with those expected for shock-dominated gas. Altogether, none of these methods provide significant evidence for shocks in the six LEGGOS sources. We conclude by discussing the implications of the lack of evidence for shocks in the broader context of galaxy evolution at cosmic noon and earlier epochs.
\end{abstract}


\section{Introduction}\label{sec:intro}
Localized increases in the pressure of the interstellar medium (ISM) can cause an irreversible fluid-dynamical disturbance (a ``shock'') which propagates into the surrounding gas \citep[for a theoretical review, see, e.g.,][]{Draine1993}. There exist several phenomena which can create shocks on galactic and sub-galactic scales, including the turbulent motions of clouds in the ISM \citep[e.g.,][]{Peimbert1991}, outflows or bubbles of relativistic electrons from stellar evolution \citep[e.g.,][]{Rich2014,Krabbe2014,Jaskot2016} or active galactic nuclei \citep[AGN; e.g.,][]{Osterbrock1971,Daltabuit1972,Heckman1980}, or mergers of galaxies or galaxy clusters \citep[e.g.,][]{Marketvitch2002,Marketvitch2007,Ha2018}. Shocks induced by any of the above phenomena can have a significant contribution to the resulting observed spectrum of a given system \citep[e.g.,][]{Dopita1995,Dopita1996,Dopita.2017,Sutherland.2017,Sutherland.2018}. As each of these phenomena are known to evolve with redshift \citep[e.g.,][]{Hopkins2007,Madau2014,Pearson2019}, the identification of shocks in high-z systems is critical to interpret their observed spectra.

A commonly used method of determining the presence of shocked gas and differentiating it from other sources in a system is through combining emission line ratios in two dimensional diagnostics. These diagnostics often leverage one emission line ratio that is sensitive to the ionization parameter, defined as the ratio of the local ionizing photon flux to the local hydrogen density \citep[e.g.,][]{Kewley2019}. The most commonly used of these ionization parameter sensitive line ratios are $\oiii~\lambda5008/\oii~\lambda\lambda3727,3730$ (hereafter O32), $\oiii~\lambda 5008/\hb$ (hereafter \oiii/\hb), $\neiii~\lambda3870/\oii~\lambda\lambda3727,3730$ (hereafter \neiii/\oii), and $\siii~\lambda9071,9533/\sii~\lambda\lambda6718,3733$ (hereafter S32). Each of these emission line ratios is comprised of a line sensitive to higher ionization energies (35.12 eV for \oiii, 40.96 eV for \neiii, and 23.34 eV for \siii) over a ``low-ionization'' line (13.60 eV for \hb, 13.62 eV \oii, and 10.36 eV for \sii). Each line ratio is chosen to be relatively insensitive to variations in abundance patterns, thus changes in the ratio can be attributed to variations in the hardness of the ionizing spectrum or the geometry of the surrounding gas \citep[e.g.,][]{Aller1942,Baldwin1981,Veilleux1987,Levesque2014,Kewley2002,Kobulnicky2004,Kewley2006,Kewley2019,Berg2021}. These ionization parameter-sensitive line ratios are then compared to line ratios more sensitive to other nebular conditions, e.g., abundances \citep[e.g.,][]{Baldwin1981,Veilleux1987,Kewley2019}. The most commonly used line ratios of this kind are $\nii~\lambda6585/\ha$ (hereafter $\nii/\ha$), $\sii~\lambda\lambda6718,6733/\ha$ (hereafter $\sii/\ha$), and $\oi~\lambda6302/\ha$ (hereafter $\oi/\ha$). 

These two dimensional emission line ratio diagrams are often divided into regions dominated by different ionizing sources, e.g., stellar populations, active galactic nuclei (AGN) and often shock-dominated low-ionization (nuclear/narrow) emission regions \citep[LI(N)ERs; e.g.,][]{Kewley2006}. Shock models suggest that shocks are capable of reproducing emission lines across a wide dynamic range in these diagnostics,  and that the \oi/\ha\ ratio in particular may hold the greatest leverage in differentiating shocks from stellar populations and AGN as a tracer of the partially ionized gas in post-shock regions \citep[e.g.,][]{Allen.2008,Kewley2013,Kewley2019}. 

Shock velocity has a significant impact on the observed emission lines; slow shocks ($v_\mathrm{shock} \lesssim500$ km s$^{-1}$) can produce line ratios similar to stellar populations in H II regions, and fast shocks ($v_\mathrm{shock} \gtrsim500$ km s$^{-1}$) can reproduce line ratios similar to AGN narrow line regions. The hot, shocked gas from a fast shock can produce photons which irradiate the preshocked gas beyond the shock front, resulting in a pre-heated and pre-ionized zone (a radiative ``precursor'') \citep[e.g.,][]{Draine1993,Allen.2008}. Contemporary shock modeling software \citep[e.g.,][]{Sutherland.2017,Dopita.2017,Sutherland.2018} can self-consistently model fast shocks and their precursors, where previous generations of models required separate handling of each component. The precursor results in stronger high-ionization emission lines \citep[$\gtrsim 35$ eV, e.g., ][]{Berg2021,Olivier2022}, thus higher ionization parameter-sensitive line ratios (e.g., \oiii/\hb, \neiii/\oii, O32). As such, shocks of varying velocities can produce emission lines attributed to other sources of ionization (e.g., stellar populations and AGN photoionization) in the aforementioned emission line ratio diagnostics. 

Following \cite{Kewley2019}, shocks can be unambiguously identified with spatially-resolved spectroscopy by leveraging kinematic information via the following criteria: (1) the velocity dispersion distribution is bimodal (2) the velocity dispersion of the broad component is $>80$ km/s (3) the velocity dispersions correlate with shock-sensitive emission line ratios (4) The line ratios are consistent with predictions from shock models. This method is limited to observations with high spectral resolution (30-50 km s$^{-1}$ at \ha) and those with depths sufficient to detect an underlying broad component. Using the spatial information from integral field spectroscopy, we can disambiguate shocks from AGN photoionization, where regions with elevated emission line ratios (often $\log\sii/\ha\gtrsim-0.5$\ and $\log\oi/\ha\gtrsim-1$) are consistent with shocks at large radii and AGN photoionization at small radii. Leveraging all or a subset of these criteria, several works have combined emission line ratios, velocity dispersions, distance from galactic centers or other energy sources, and other spectral and morphological information to separate shocks, AGN, and stellar populations with spatially resolved spectroscopy for local galaxies \citep[e.g.,][]{Ho2014,Belfiore2016,DAgostino2018,DAgostino2019,Law2021,Long2022,Alban2023,Zhu2025}. Similar methods have been applied in case studies to trace shocks at $z\sim1$ \citep[e.g.,][]{Yuan2012}, though studies at higher redshifts have not been possible before current instrumentation.  

The introduction of JWST \citep{Gardner2006,Gardner2023} and the Near-Infrared Spectrograph \citep[NIRSpec;][]{Jakobsen2022} allows for unprecedented spectroscopy of the full suite of rest-frame optical emission lines to $z
\lesssim 6$. Additionally, integral field spectroscopic mode on the NIRSpec instrument \citep[IFS,][]{boker2022} enables for the first time spatially-resolved spectroscopic disambiguation of shocked gas from AGN and stellar photoionization at these epochs. To date, the search for shocks at high redshifts with JWST has been limited to case studies or small samples of integrated galaxy spectra \citep[e.g.,][]{Flury2025,DEugenio2025}. Here we present the first systematic search for shocks in gravitationally-lensed galaxies at $z\sim2-4$ with spatially-resolved spectroscopy from JWST/NIRSpec.

The remainder of this work is as follows. In Section \ref{sec:data}, we introduce the JWST integral field spectroscopy from the LEGGOS Survey. In Section \ref{sec:models}, we describe the shock and photoionization models which we compare to the observations. In Section \ref{sec:results}, we report our findings on the presence of shocks in the LEGGOS galaxies. In Section \ref{sec:discussion}, we discuss the implications of shocked gas in the interpretation of spatially-resolved spectra at these epochs. Finally, in Section \ref{sec:summary}, we summarize our conclusions and indicate future pathways for this work.

\section{Data}\label{sec:data}

We leverage JWST/NIRSpec integral field spectroscopy \citep{boker2022} from the LEnsing and Galaxy Growth: Observing Substructures (LEGGOS) Survey \citep[][GO-4125 \& GO-3843, PI: Khullar, Florian, Bayliss]{Khullar2026}, which include six new IFS observations of gravitationally-lensed galaxies, and two archival sources pulled from previous NIRSpec/IFS data (Sunburst Arc, GO-2555, PI: Rivera-Thorsen; SGAS1226, from the TEMPLATES program, GO-1355, PI: Rigby, Vieira). For additional details on these archival programs, see \cite{riverathorsen2025arxiv} and \cite{Rigby2025}, respectively.
The observations were designed to simultaneously sample a large spectral range in rest-wavelength space -- covering all of the strong rest-frame optical emission lines commonly used in high-redshift galaxy science \citep[e.g.,][]{Kewley2019} in the medium-resolution grating -- as well as reach depths required to measure the stellar continuum in each source using the prism.  The IFS observations are described in detail in \citet{Khullar2026}, but we summarize the main points and reduction information here.

In this work we analyze six of the eight lensed galaxies from the LEGGOS survey with spectroscopic redshifts between $z=2.37{-}3.63$.  We did not fully analyze two sources, SGAS2111 and SGAS1226, due to lack of detectable fainter emission lines needed in this work, though we briefly comment on SGAS2111 in Section \ref{sec:discussion}.  The resulting six galaxies and IFS data used are summarized in Table \ref{tab:sources}. 
The Sunburst Arc data include three total pointings, with two of those (dubbed P2 and P3, here referred to as P2P3) overlapping on the sky, while the third (dubbed P1) is offset by $\sim 12$\arcsec \citep[see][for additional details]{riverathorsen2025arxiv}. We divide the Sunburst Arc into two ``sources" in Table \ref{tab:sources} and in subsequent figures for visual clarity, however we note that these pointings cover some of the same physical regions in the source galaxy; for example the Lyman-continuum leaker \citep{riverathorsen2019} appears multiple times in both P1 and P2P3 \citep{riverathorsen2024}. We considered adding three additional targets from the TEMPLATES program \citep{Rigby2025} but elected not to include these due to a lack of coverage or detection of the key \oi$\lambda6302$ emission line \citep{welch2024}.


The full data reduction process for the NIRSpec data is described in \cite{Khullar2026}. Here we briefly summarize the key steps. We used the JWST data reduction pipeline version 1.20.2, with the corresponding calibration reference data set (CRDS) \texttt{jwst1466.pmap}. We use the \texttt{clean\_flicker\_noise} step in the first stage of the pipeline to remove 1/f noise from the detector images, selecting the \texttt{fft} parameter to use the NSClean algorithm \citep{Rauscher24_NSClean}. We use the native 0\farcs1 spaxel size to minimize the effects of the undersampled NIRSpec detector pixels \citep{Law23drizzle}. 

After the standard data reduction pipeline, we apply two post-processing steps. First, we use the \texttt{baryon-sweep} code to remove any remaining outliers and artifacts that were not caught by the standard pipeline outlier detection step \citep{Hutchison24baryonsweep}. Second, we subtract the predicted background spectrum from the JWST Background Tool (JBT)\footnote{\href{https://jwst-docs.stsci.edu/jwst-other-tools/jwst-backgrounds-tool}{https://jwst-docs.stsci.edu/jwst-other-tools/jwst-backgrounds-tool}}. Previous works have shown that the predicted backgrounds are sufficient for medium and high resolution NIRSpec IFU observations \citep[e.g.,][]{Rigby2025,Khullar2026}.

\begin{deluxetable*}{lccccc}[!ht]
\tablecaption{\label{tab:sources}NIRSpec/IFS observations in this work, from the broader LEGGOS Survey}
\tablecolumns{6}
\tabletypesize{\normalsize}
\tablewidth{\textwidth}
\tablehead{
\colhead{Galaxy Name} & \colhead{RA [deg]} & \colhead{Dec [deg]} & \colhead{$z_{spec}$} & \colhead{Disperser/Grating} & \colhead{$t_{exp}$ [s]}
}
\startdata
SGAS1050 & 162.6641 & $+$0.2915 & 3.625 & G235M/F170LP, prism/clear & 8870, 5952  \\
Cosmic Eye & 323.8029 & $-$1.0286 & 3.074 & G235M/F170LP, prism/clear & 8870, 5952  \\
SGAS1429 & 317.8280 & $-$1.2415 & 2.824 & G235M/F170LP, prism/clear & 8870, 5952  \\
SGAS1527 & 217.4788 & $+$12.0439 & 2.762 & G235M/F170LP, prism/clear & 8870, 5952  \\
SGAS1110 & 231.9389 & $+$6.8720 & 2.4795 & G235M/F170LP, prism/clear & 8870, 5952  \\
Sunburst Arc P1 & 167.5831 & $+$64.9978 & 2.3709 & G140H/F100LP, G235H/F170LP & 5894, 8870  \\
Sunburst Arc P2P3  & 237.5183 & $-$78.1833 & 2.3709 & G140H/F100LP, G235H/F170LP & 5894, 8870  \\
\enddata
\end{deluxetable*}

\subsection{Emission Line Fits and Flux Measurements}

For each source, we fit a single gaussian to each detected emission line.  The observed line width of each emission feature was described by the combination of the intrinsic line width and the instrumental resolution at the observed wavelength added in quadrature -- with the instrumental resolution for JWST/NIRSpec IFS adapted from \citet[for the medium and high resolution gratings]{Shajib.2025} and O'Brien et al.,\ in prep (for prism). For line complexes such as \hb\ + \oiii\ \lam4960,5008; \nii\ \lam6550,6585 + \ha; and \sii\ \lam6717,6732, we fit all lines in a complex together such that width and relative strengths of each line are all connected \citep[for more details, see][]{hutchison2026,welch2025}.  The fitting routines were run on both the individual spaxels and the spatially-integrated spectra (see \ref{subsec:int-spec}) for the data of each source.  In total, the lines fit for each source in this work include those used in canonical BPT-like line-ratio diagnostics.

For two sources -- namely SGAS1050 and both pointings of Sunburst Arc -- we detect faint broad components in multiple lines for a small number of spaxels.  For these sources + spaxels, we fit two components to each line, with the widths and velocity offsets from systemic for all broad components fixed to each other. We note that while broad components may be detected in multiple lines for some spaxels, we did not detect a broad component in the faint \oi\ \lam6302 emission line for any of these spaxels.

In the rest of this work, we report only the line fluxes and measurements from single-gaussian fits to each emission line.  However, when a broad component is significantly detected ($S/N \geq 3$) in each line used in a diagnostic, we include the resulting broad component measurements in the analysis detailed in this work.

\subsection{Spatially-Integrated Spectroscopy} \label{subsec:int-spec}

In addition to the full spatially-resolved (i.e. spaxel-by-spaxel) measurements of the various spectral features used in this work, we also measure spatially-integrated properties to visualize the differences between the full galaxy and its smaller sub-regions. The spatially integrated spectra also improve the detectability of fainter features, particularly the faint \oi$\lambda6302$ line.

\subsubsection{Full-arc integrated spectra}

For five of the six galaxies considered in this work, we define the full-arc integrated spectra using the continuum-subtracted S/N of the \ha\ emission line in the G235M observations. The full-arc spectra include all spaxels with $\text{S/N}(H\alpha) > 5$. A thorough description of the full-arc spectral extraction is provided in Welch et al., in prep. Here we provide a brief overview of the process. 

We measure the continuum-subtracted S/N in each spaxel by summing the flux density in a wavelength region around the \ha\ line, and separately summing the flux density in an identically sized wavelength region with no detectable emission lines around $\sim 6450$\AA. For the G235M spectra, we use a region 6 resolution elements wide for both the emission line and continuum measurements. We then subtract the continuum flux from the line flux, and add the uncertainties from each region in quadrature. 

The final full-arc spectrum includes the summed flux from all spaxels with continuum-subtracted $\text{S/N} > 5$, with uncertainties summed in quadrature. These spectra are not corrected for lensing magnification, though the magnification does not affect emission line ratios so the magnification correction is not critical for this work. The resulting full-arc spectra are displayed in black in Figure \ref{fig:spectra}.

We use a different procedure to extract the full-arc spectrum of the Sunburst Arc because several bright regions are imaged many times and would likely bias the full-arc spectrum. We use the galaxy-integrated spectrum described in \cite{riverathorsen2025arxiv}, which uses a custom mask to define a single image of the galaxy, thereby avoiding biases from multiple images. The Suburst spectrum from \cite{riverathorsen2025arxiv} also includes a magnification correction, but again this does not affect emission line ratios and therefore does not affect the results of this analysis.

\subsubsection{Clump Spectra} \label{subsubsec:clumps}

\begin{figure*}
    \centering
    \includegraphics[width=0.99\linewidth]{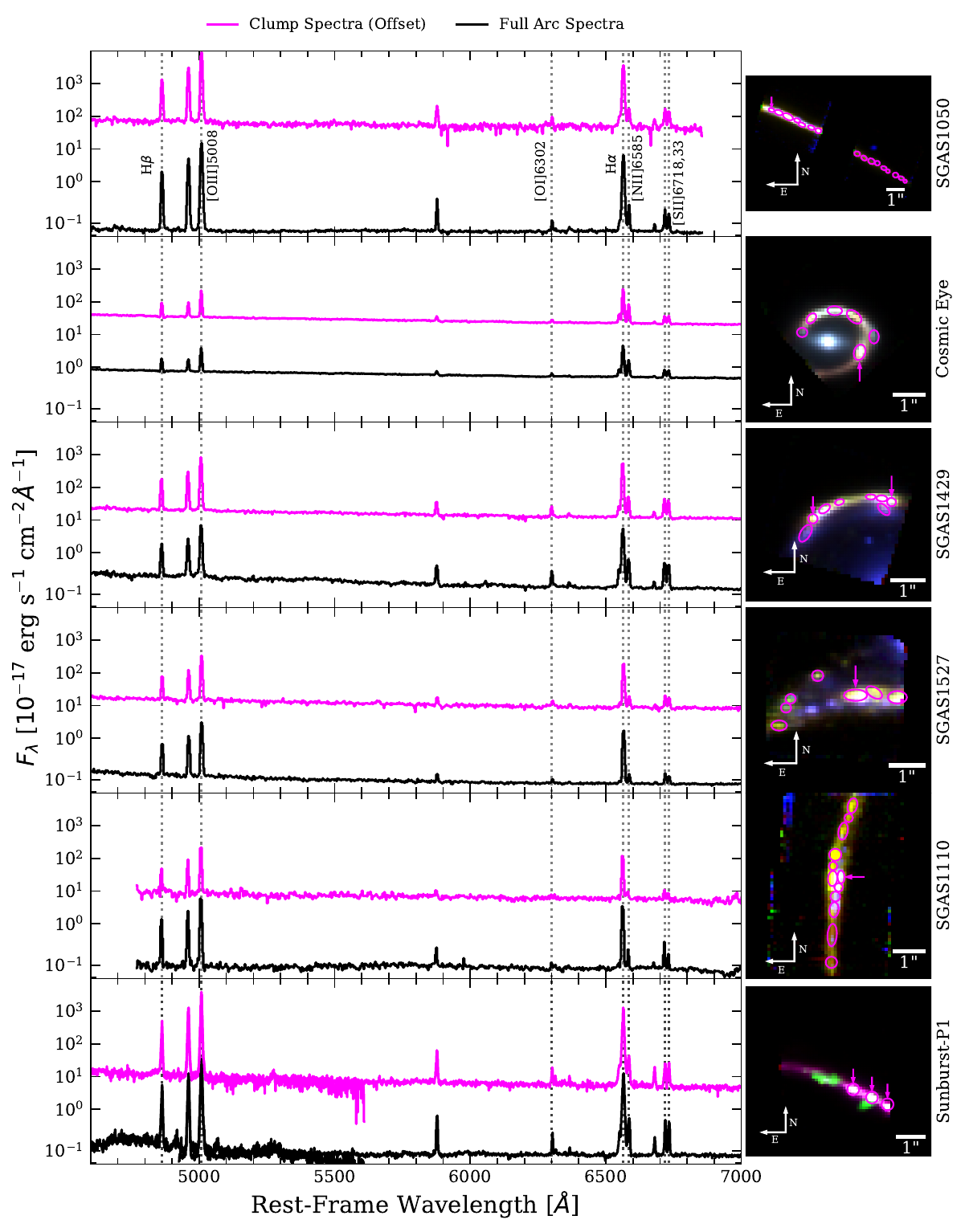}
    \caption{Representative spectra for each of the LEGGOS targets considered in this work. The black line in each panel shows the full-arc integrated spectrum, and the pink line shows a representative clump spectrum. The clump spectra are shifted with a multiplicative offset for visual clarity. The color images on the right of each row show NIRSpec/IFS data for each target, with H$\alpha$ emission in red, \oiii$\lambda5008$\ emission in green, and stellar continuum emission in blue. Clump regions are shown as pink ellipses in the color images, and pink arrows indicate which clump spectra is plotted. Multiple arrows indicate multiple images which have been stacked. }
    \label{fig:spectra}
\end{figure*}

As part of this work, we identify star-forming clumps in each source in order to improve faint line detections while retaining some spatial information. We identify clumps visually using a combination of spatially-resolved \ha\ and continuum flux maps (see also Welch et al., in prep). The initial clump identification utilizes the \ha\ maps, and stellar continuum flux only contributes in cases where large regions of the lensed arc show fairly smooth \ha\ emission, most prominently in the brightest region of SGAS1050. Identified clumps are circled in the color images in the right-hand panels of Figure \ref{fig:spectra}. 

Identifying clumps based on \ha\ emission preferentially selects HII regions powered by recent star formation. This is advantageous since we are primarily interested in nebular emission lines originating in such HII regions. We note that this identification will necessarily differ from clumps identified for other purposes, e.g. gravitational lens modeling \citep[e.g.,][]{sharon2020,Sharon2022sunburst,Abedi2026} or photometric analysis \citep{Khullar2026}. 

For the Sunburst Arc, the only clump we analyze is the Lyman-continuum leaker, which has been extensively studied previously using NIRSpec data \citep[e.g.,][]{riverathorsen2024,welch2025}. We do not define additional clumps for this target since the S/N is high enough across the full arc to provide sufficient detail for our analysis.

\subsection{Comparison to MaNGA ``eLIER'' 1-550578} \label{sec:mangadata}
For a shock-dominated, spatially-resolved anchor at low redshift in the following analysis, we compare to the LI(N)ER 1-550578 (RA=324.833648832, Dec=10.54076696, $z=0.0766$; \citealt{Belfiore2016}). 1-550578 has integral field spectroscopy from the Sloan Digital Sky Survey-IV \citep[SDSS-IV;][]{Blanton2017} Mapping Nearby Galaxies at Apache Point Observatory \citep[MaNGA;][]{Bundy2015} survey. We access the data for 1-550578 using the SDSS Marvin application \citep{brian_cherinka_2017_292632,Cherinka2019}.

1-550578 is identified as a shock-dominated ``extended LIER'' (denoted eLIER in the following analysis) by \cite{Belfiore2016}, and has \oiii/\hb\ and \sii/\ha\ line ratios consistent with the LI(N)ER regions of the \cite{Kewley2006} diagnostics for nearly all (82\% for the \oiii/\hb\ versus \sii/\ha\ diagnostic) spaxels within the MaNGA field of view. We select spectra from 1-550578 with signal to noise ratio $S/N>5$ for all emission lines in Figure \ref{fig:ratios_leggos_part1}. 

\section{Models}\label{sec:models}

The shock models used in this work are from the Mexican Million Models database\footnote{The 3MdB$^S$ shock model database and documentation can be accessed here: \href{http://3mdb.astro.unam.mx:3686}{http://3mdb.astro.unam.mx:3686}} \citep[3MdB$^S$;][]{Morisset.2015,Alarie2019}. 3MdB$^S$ has a suite of $\sim2\times10^5$ total shock models computed with MAPPINGS V v5.1.13 \citep{Sutherland.2017,Dopita.2017,Sutherland.2018}. The modern iterations of MAPPINGS self-consistently model fast shocks and their photoionizing precursors such that they need not be modeled distinctly. 

3MdB$^S$ includes shock models of varying metallicities and abundance patterns, pre-shock densities, shock velocities, and magnetic field strengths \citep{Alarie2019}. For this work, we use the shock only and shock plus precursor grids corresponding to the \cite{Allen.2008} solar abundance pattern (see the top row of Figure \ref{fig:ratios_leggos_part1}); for a complete discussion on the effects of abundances on shock model outputs, see Section 3 of \cite{Allen.2008} and Section 4 of \cite{Alarie2019}. The LEGGOS sources span a wide range of gas-phase metallicities \citep[see, e.g.,][as well as B. Welch et al. in preparation]{Khullar2026}. The models are computed over grids of pre-shock density $-2 \leq \log n/\mathrm{[cm^{-3}]} \leq 3$ in steps of $\Delta\log n/\mathrm{cm^{-3}} = 1$, shock velocity $100\leq v_\mathrm{shock}/\mathrm{km~s^{-1}}\leq1000$ in steps of $\Delta v_\mathrm{shock}/\mathrm{km~s^{-1}} = 25$, and a magnetic field grid which is non-uniform in $B$ and dependent on the pre-shock density with a minimum $B=0.0001$ $\mu$G and maximum $B=1000$ $\mu$G (see the online documentation and \cite{Alarie2019} for further details). We discuss the limitations of the grid and the inference of the shock parameters in Section \ref{sec:discussion:improvements}.

\section{Results}\label{sec:results} 

In this Section, we employ several different selection methods to search the LEGGOS sample for signatures of shocked gas. We begin this exploration with the standard rest-frame optical emission line ratio diagnostics from \cite{Kewley2006}. 

\begin{figure*} 
    \centering
    \includegraphics[width=0.9\linewidth]{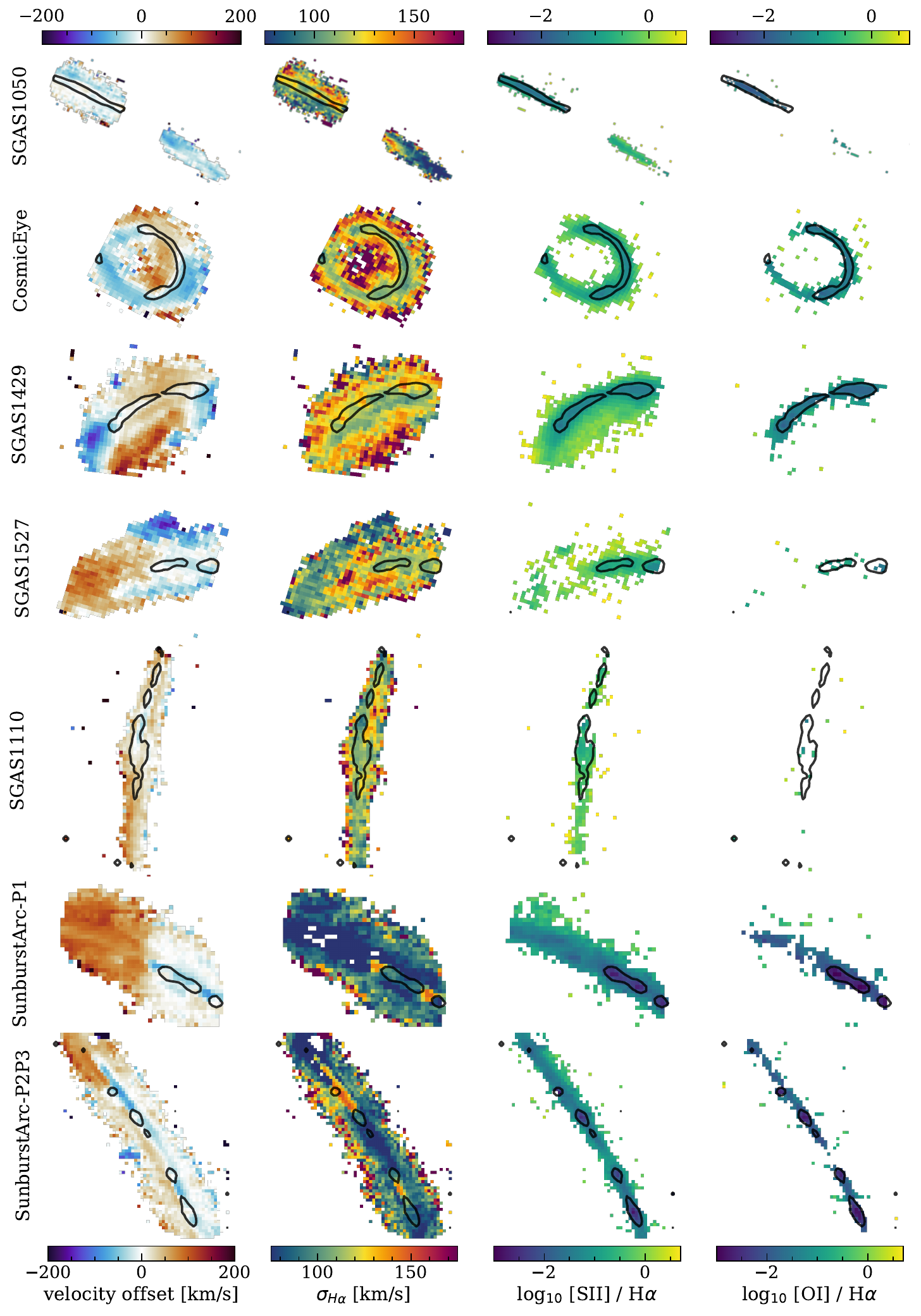}
    \caption{Image-plane maps of the LEGGOS sources explored in this work showing (left) the velocity offset from systemic, (center left) the velocity dispersion of the narrow component of nebular emission, and (center right and right) the \sii/\ha\ and \oi/\ha\ flux ratio diagnostics, respectively, frequently used to identify regions of partially-ionized gas which may be excited by shocks. All panels are oriented north up, east left, and the black contours in each panel denote regions with strong \ha\ emission. }
\label{fig:image_plane_maps}
\end{figure*}

Figure \ref{fig:image_plane_maps} shows image plane maps of the shock-sensitive line ratios \sii/\ha\ and \oi/\ha, alongside the velocity offset and velocity dispersion measured from the \ha\ emission line. We note that most targets have more complete coverage of \sii/\ha\ than \oi/\ha\ due to the faintness of the \oi$\lambda6302$ line. We utilize these line ratios and kinematics in the following analysis.

\subsection{Shock-Sensitive Line Ratio Diagnostics}

Figures \ref{fig:ratios_leggos_part1} and \ref{fig:ratios_leggos_part2} shows the emission line ratio diagnostics \oiii/\hb\ versus \nii/\ha, \sii/\ha, and \oi/\ha\ for the MAPPINGS-V shock plus precursor models from 3MdB along with the integrated, clump, and individual pixel spectroscopy for the LEGGOS sources. Using the \cite{Kewley2006} methodology with demarcation lines from \cite{Kewley2001} and \cite{Kauffmann2003}, we find that the LEGGOS sources are maximally $\sim8$\% (by unweighted pixels in the image plane) identified as shocks by these diagrams. We tabulate the image plane shock pixel fractions in Table \ref{tab:bptresults}. The sources with the highest shock percentages in \oi/\ha\ are also those with the least coverage in the \oi\ emission line. SGAS1527 has the highest percentage of shock-consistent spaxels in the \oi/\ha\ ratio  at 7.89\%, but only 38 total spaxels meet the S/N(\oi)$>2$ criterion we require to measure this ratio. The second-highest percentage in the \oi/\ha\ ratio is from SGAS1110 at 6.67\%, however here the sample size is even smaller at only 15 spaxels meeting the S/N requirement. Of the objects with at least 100 spaxels meeting the S/N requirement, the highest fraction of shock-consistent spaxels is 3.23\% across either the \oi/\ha\ or \sii/\ha\ diagnostics. Additionally, none of the shock-identified pixels show any coherent structure in the image plane maps shown in Figures \ref{fig:ratios_leggos_part1} and \ref{fig:ratios_leggos_part2}, further indicating that they may be driven by statistical scatter. 

\begin{deluxetable}{lccccc}[!ht]
\tablecaption{\label{tab:bptresults}Shock Percentages in LEGGOS Galaxies}
\tablecolumns{3}
\tabletypesize{\normalsize}
\tablewidth{\textwidth}
\tablehead{
\colhead{Galaxy Name} & \colhead{\sii/\ha} & \colhead{\oi/\ha} 
}
\startdata
SGAS1050 & 2.3\% (9/391) & 0.92\% (1/109)   \\
Cosmic Eye & 1.67\% (7/419) & 2.93\% (6/205) \\
SGAS1429 & 1.25\% (6/481) & 0\% (0/169)  \\
SGAS1527 & 0.33\% (1/305) & 7.89\% (3/38) \\
SGAS1110 & 1.25\% (3/240) & 6.67\% (1/15)  \\
Sunburst Arc P1 & 0\% (0/402) & 3.23\% (5/155)  \\
Sunburst Arc P2P3  & 0\% (0/551) & 1.7\% (4/235)   \\
\enddata
\tablecomments{Percentages of individual spaxel spectra that meet the LI(N)ER criteria from \cite{Kewley2001,Kewley2006} diagnostic demarcations, possibly indicating the presence of shocks. The ratios in parentheses are the total number of LI(N)ER spaxels over the total number of spaxels with S/N$>2$ in each respective emission line. We do not tabulate the \nii/\ha\ shock percentages as the \cite{Kewley2006} demarcations for the \oiii/\hb\ versus \nii/\ha\ diagram do not have an explicit LI(N)ER region.}
\end{deluxetable}

We next explore the relationship between the kinematics and the observed emission line ratios. Figures \ref{fig:velocity_dispersion_line_ratios_part1} and \ref{fig:velocity_dispersion_line_ratios_part2} show the velocity dispersion derived from \ha, $\sigma_{H\alpha}$, versus three of the emission line ratios from Figures \ref{fig:ratios_leggos_part1} and \ref{fig:ratios_leggos_part2}: \nii/\ha, \sii/\ha, and \oi/\ha. The vertical lines in the center and right panels indicate the lower limit for the shock regions in their respective diagnostics from Figures \ref{fig:ratios_leggos_part1} and \ref{fig:ratios_leggos_part2}. Previous works have noted a correlation between the $\sigma_{H\alpha}$ and these line ratios for shocked gas \citep[e.g.,][]{RodriguezdelPino2019,Vayner2023}. For both the single component and broad component fits, we do not observe such a correlation in the shock regime of each line ratio.


\subsection{Comparison with Shock Models}

We now explore the relationships between physical quantities derived from the spectra and quantities inferred from the 3MdB$^S$ MAPPINGS-V shock models. We infer shock velocities from the observed emission line fluxes and their uncertainties for each emission line in Figures \ref{fig:ratios_leggos_part1} and \ref{fig:ratios_leggos_part2}, omitting emission lines without fits for individual spectra. We calculate likelihoods by their $\chi^2$ distances from the models, i.e., $\ln \mathcal{L} = \frac{1}{2}\sum_i(\mathrm{data_i} - \mathrm{model})^2/\mathrm{uncertainty_i}^2$. We compute the $\chi^2$ normalization for each model analytically \citep[see Appendix A of ][]{Sawicki2012}. We apply the  uniform prior for shock velocity set by the 3MdB$^S$ shock model grid, i.e., $v_{\text{shock}}/\mathrm{km s}^{-1}\sim U[100,1000]$ (see Section \ref{sec:models}).

\begin{figure*}[ht]
\centering
\epsscale{1.15}
\plotone{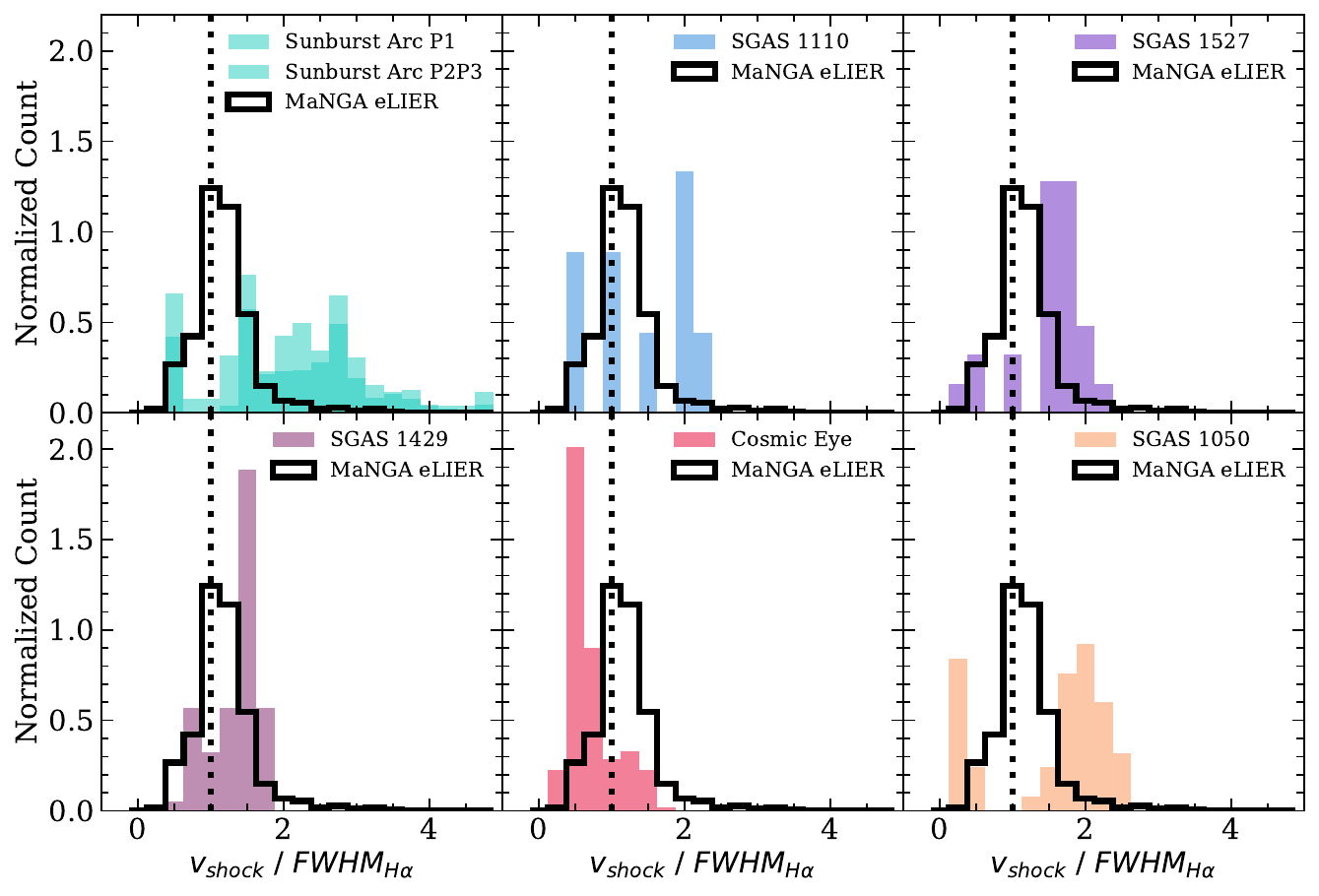}
\caption{Histogram showing the ratio of the inferred shock velocities to the full width at half-maximum (FWHM) of \ha\ for the LEGGOS sources compared to the MaNGA ``eLIER'' 1-550578. For shocked gas, the ratio $v_{shock}$/FWHM$_{H\alpha}$ should scale as the geometric factor $\epsilon$ (See Section \ref{sec:results}). We find that all LEGGOS galaxies with at least ten spaxels with measured line fluxes and FWHM$_{H\alpha}$ follow significantly different distributions from the MaNGA eLIER 1-550578, suggesting that they are not well described by the parameters inferred from the shock models. 
\label{fig:vshock_fwhm_hist_six_panel}} 
\end{figure*}

We use shock models to test for shocked gas in spectra which otherwise are not identified as shocks from Figures \ref{fig:ratios_leggos_part1} and \ref{fig:ratios_leggos_part2}. The inferred shock velocities are nominally correlated with the velocity dispersions measured from the widths of observed emission lines. The observed velocity dispersion is dependent on the geometry of the shock, where it will be maximized for shocks moving in opposite directions along the line of sight, and minimized for shocks traveling transversely to the line of sight \citep[e.g.,][]{Dopita2012}. Thus, the shock velocity will, to first order, be proportional to the velocity dispersion with some geometric coefficient $\epsilon$. This allows us to test for shocked gas particularly in the slow shock regime, where shocked gas correlates linearly with the velocity dispersion with slope $\epsilon$. We explore the potential for this relation in Figure \ref{fig:vshock_fwhm_hist_six_panel}. We show that the distributions of $v_{\text{shock}}/\text{FWHM}_{\ha}$ are significantly different for the shock-dominated MaNGA source 1-550578 and all but one of the LEGGOS sources, where 1-550578 is centered around unity and approximately Gaussian distributed. We employ a two-sample Kolmogorov-Smirnov test \citep{Karson1968} for each of the LEGGOS inferred $v_{shock}/\text{FWHM}_{\ha}$ compared to the MaNGA eLIER 1-550578; we reject the null hypothesis that the samples are drawn from the same parent distribution at a $p<0.05$ significance for all LEGGOS galaxies except for SGAS1110, which returns a marginal $p=0.082$. We attribute this to small-number statistics as SGAS1110, which has an order of magnitude fewer pixels with all measured lines and FWHM$_{H\alpha}$ (9) than the rest of LEGGOS, driven primarily by a lack of $\oi~\lambda6302$ (see Figure \ref{fig:ratios_leggos_part2}).

\subsection{Merger-Induced Shocks}
\begin{figure*}
    \centering
    \includegraphics[width=\linewidth]{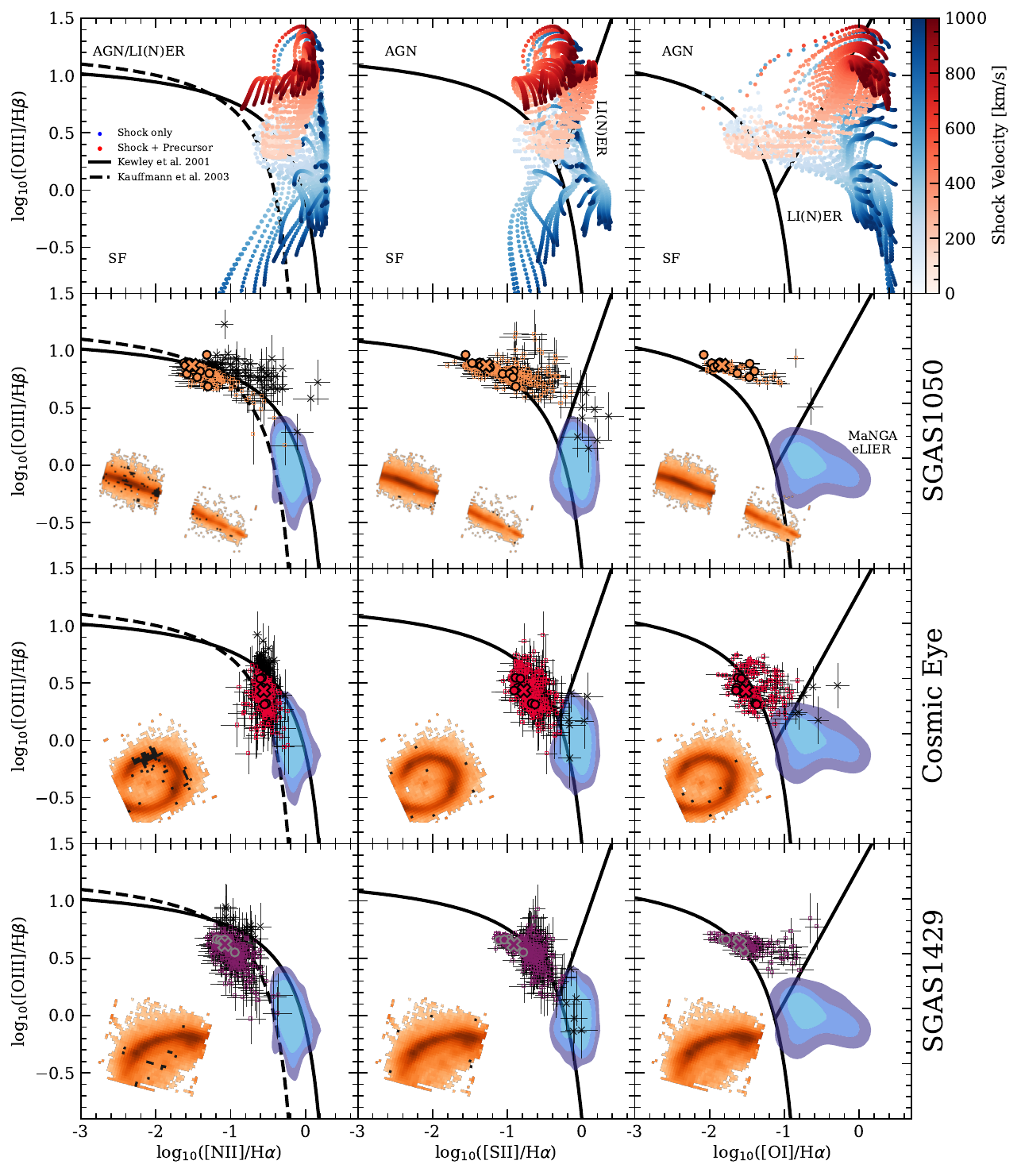}
    \caption{The \oiii/\hb\ versus \nii/\ha\ (left), \sii/\ha\ (center), and \oi/\ha\ (right) line ratio diagnostics using the separations between star-forming, AGN, and LI(N)ER regions from \cite{Kewley2001} (solid black lines) and \cite{Kauffmann2003} (dashed black line). The top row of panels show shock-only models (blue) and shock-plus-precursor models (red) from 3MdB$^S$, color-coded by shock velocity. The bottom three rows show the first three LEGGOS targets. Full-arc integrated measurements are shown as an X, clump integrated measurements are shown as circles, and individual spaxels from the IFU maps are shown as squares. Spaxels lying in the LI(N)ER regions defined by the \cite{Kewley2001} and \cite{Kauffmann2003} curves are denoted with a black X. H$\alpha$ line maps are shown as orange insets, with black spaxels indicating the location of the possibly shock-like line ratios. For comparison, we show the MaNGA eLIER source (Sec. \ref{sec:mangadata}) as a blue contour. We see no clear evidence of shock-dominated regions in the LEGGOS targets based on these diagnostic line ratios. }
    \label{fig:ratios_leggos_part1}
\end{figure*}

\begin{figure*}
    \centering
    \includegraphics[width=\linewidth]{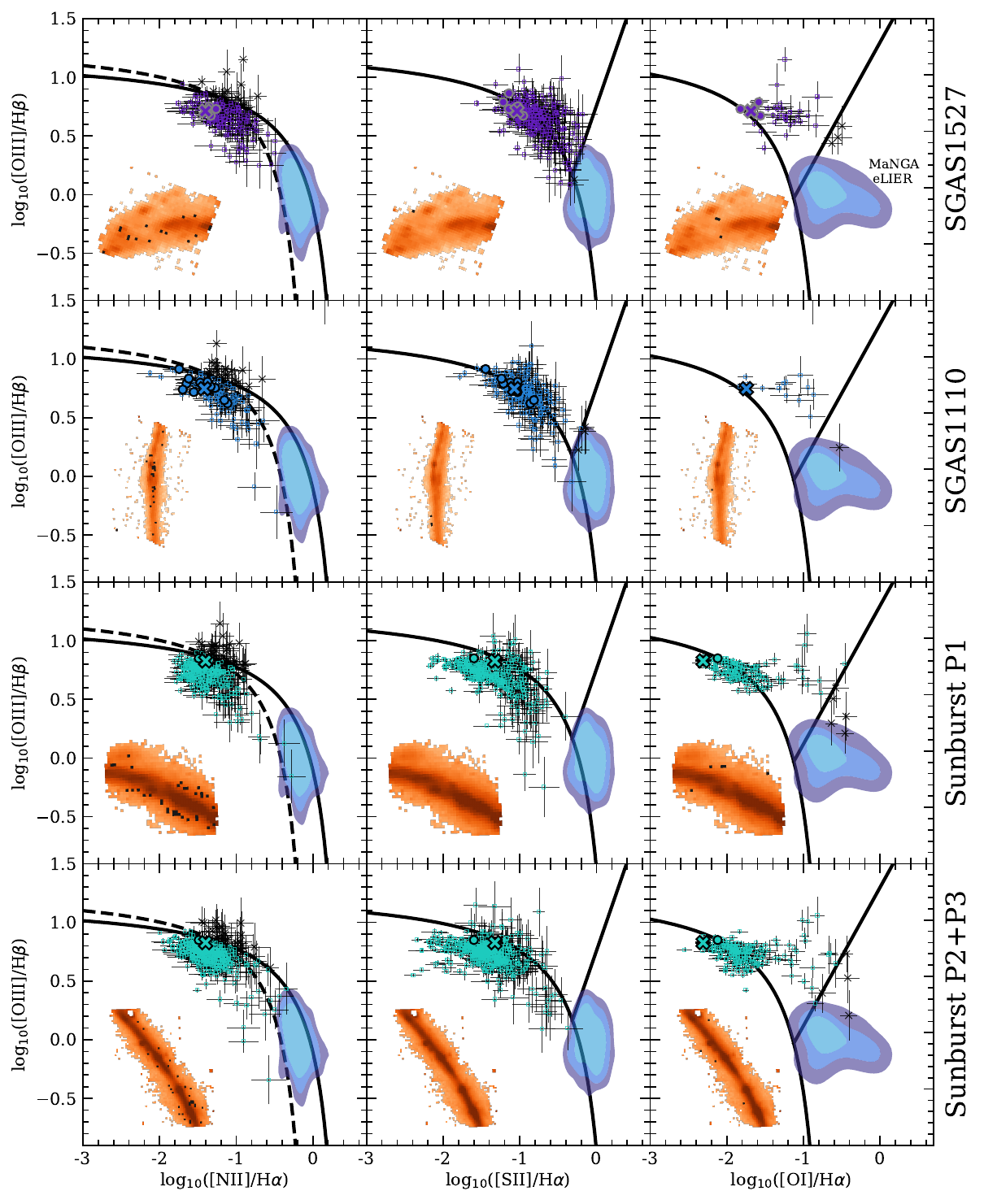}
    \caption{Same as Figure \ref{fig:ratios_leggos_part1} for the remaining LEGGOS targets. We see no clear evidence of shock-dominated regions in the LEGGOS targets based on these diagnostic line ratios.}
    \label{fig:ratios_leggos_part2}
\end{figure*}

To test for a final source of shocks induced via mergers, we look briefly at the LEGGOS spectroscopy of SGAS 2111 ($z=2.858$). SGAS 2111 is a known merging system which displays evidence of mature stellar populations with little nebular emission \citep{Khullar2026}. Galaxy mergers can produce shocks independent from stellar evolution and AGN \citep[e.g.,][]{Marketvitch2002,Marketvitch2007,Ha2018}, thus we may expect SGAS 2111 to show indications of shocks in its pixel scale spectra. We choose not to show SGAS 2111 in the above analyses as the majority of spaxels do not have detections for the emission lines shown in Figures \ref{fig:ratios_leggos_part1} and \ref{fig:ratios_leggos_part2} (fewer than 10 spaxels have significant detections of either \nii\ or \sii). Notably, SGAS 2111 shows no pixel scale detections of $\oi~\lambda6302$, which is the strongest tracer of shocks in Figures \ref{fig:ratios_leggos_part1} and \ref{fig:ratios_leggos_part2}. This represents an additional lack of evidence for shocks induced by mergers, though this is limited to a single low signal to noise source. A full analysis of the kinematics derived from the SGAS 2111 spectra is active work (R. Oh et al., in prep.).

\begin{figure*} 
    \centering
    \includegraphics[width=0.95\linewidth]{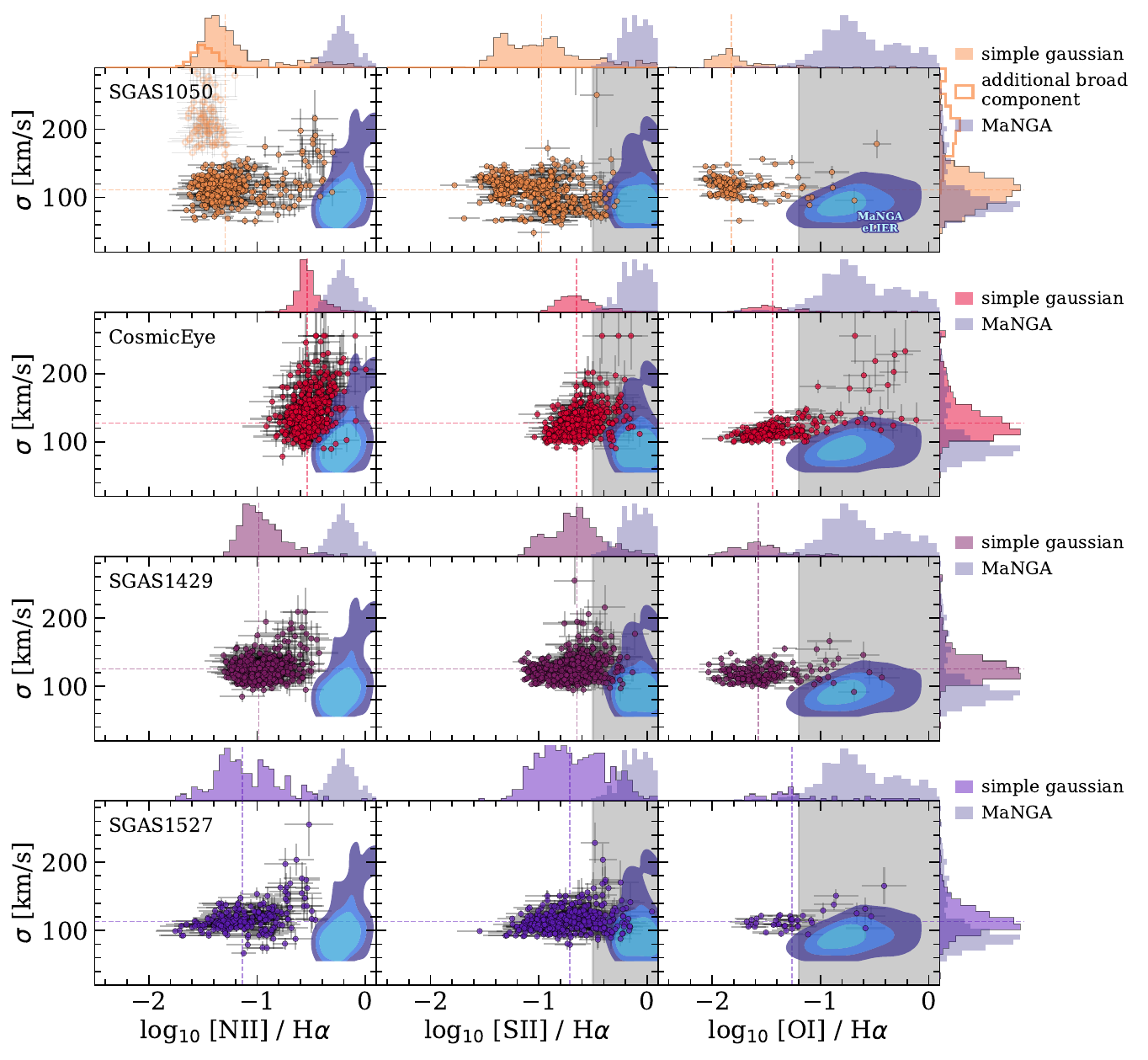}
    \caption{Velocity dispersion versus emission line ratios. Each row shows a different LEGGOS source (colored circles) compared to the MaNGA eLIER 1-550578 (blue contours). The histograms on the top and right axes of each row show the distributions in the line ratios and velocity dispersions, respectively. The gray shaded regions show the approximate shock regions for \sii/\ha\ and \oi/\ha\ from the diagnostics in Figures \ref{fig:ratios_leggos_part1} and \ref{fig:ratios_leggos_part2}. The dashed lines show the medians for the line ratios and velocity dispersions. For SGAS1050, we show the single component and broad component \nii/\ha\ line ratios as open circles. }
\label{fig:velocity_dispersion_line_ratios_part1}
\end{figure*}

\begin{figure*} 
    \centering
    \includegraphics[width=0.95\linewidth]{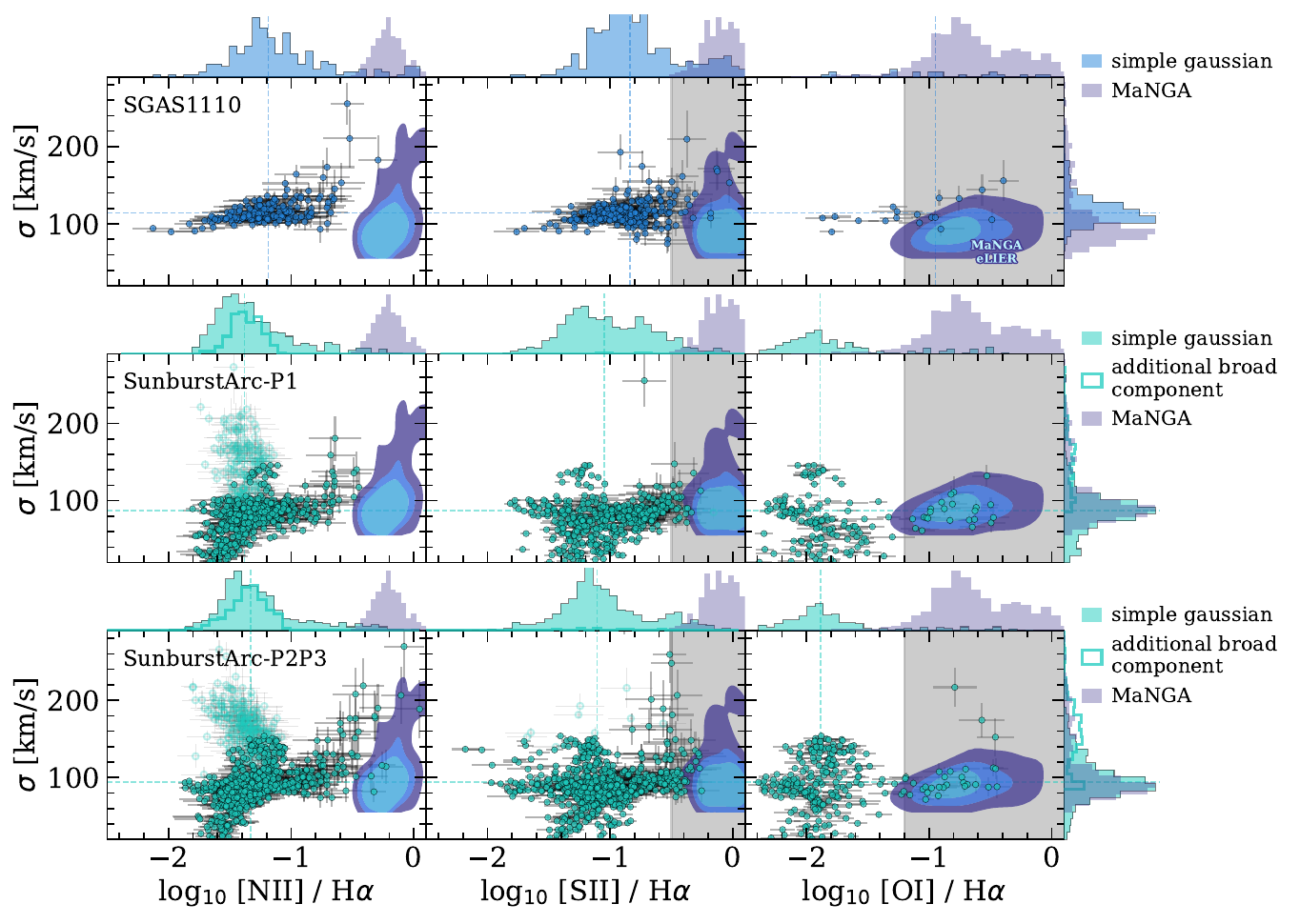}
    \caption{The same as Figure \ref{fig:velocity_dispersion_line_ratios_part1} for SGAS1110 and the Sunburst Arc. For the Sunburst Arc, we show \nii/\ha\ and \sii/\ha\ line ratios with broad component fits as open circles.}
\label{fig:velocity_dispersion_line_ratios_part2}
\end{figure*}

\section{Discussion}\label{sec:discussion}

\subsection{The Lack of Evidence for Shocks in LEGGOS}
In Section \ref{sec:results}, we present the first systematic search for shocks at $z\sim2-4$ using JWST/NIRSpec integral field spectroscopy from the LEGGOS survey. We find that shocks have minimal, if any, contribution to the observed spectra (e.g., Figures \ref{fig:ratios_leggos_part1}-\ref{fig:ratios_leggos_part2} and \ref{fig:velocity_dispersion_line_ratios_part1}-\ref{fig:velocity_dispersion_line_ratios_part2}). Here we discuss the implications of these findings.

The emission line ratio diagnostics in Figures \ref{fig:ratios_leggos_part1} and \ref{fig:ratios_leggos_part2} show that the bulk of the LEGGOS spectroscopy, be it integrated, clump-wise, or spaxel-wise, follow the standard H II region sequences which evolve as a function of ionization parameter and gas-phase abundances \citep[e.g.,][]{Kewley2013,Kewley2019,Shapley2025,Cleri2025,Cleri2026}. We note that none of the integrated or clump line ratios (larger/bolded symbols on Figures \ref{fig:ratios_leggos_part1} and \ref{fig:ratios_leggos_part2}) are consistent with the shock regions in the \oiii/\hb\ versus \sii/\ha\ or \oi/\ha\ diagrams. However, 26 and 20 spectra of individual spaxels which have line ratios consistent with shocks by the \cite{Kewley2006} diagnostics with \sii/\ha\ and \oi/\ha\ respectively, which represent up to $\sim8$\% of the spaxels in their galaxies with S/N$>2$ for the relevant emission lines. The lack of coherent structure in the shock-identified pixels in the image plane maps of Figures \ref{fig:ratios_leggos_part1} and \ref{fig:ratios_leggos_part2} suggest that these pixels may be driven by scatter. Otherwise, this result may indicate that the selection methods to identify shocks may need to evolve with redshift, similarly to those for star forming regions and AGN \citep[e.g.,][]{Kewley2013,Kewley2019,Cleri2025}.

We additionally employ the MAPPINGS-V shock models from the 3MdB$^S$ database to include another set of shock diagnostics. First, we found no relation between the inferred shock velocities and the velocity dispersions derived from \ha\ (Figure \ref{fig:vshock_fwhm_hist_six_panel}). We would expect for shocked gas that this would follow $v_{\text{shock}} = \epsilon\sigma$, where $\epsilon$ is the geometric factor as discussed in Section \ref{sec:results}. We find that $v_{\text{shock}}/\text{FWHM}_{\ha}$ is approximately Gaussian and centered near unity for the MaNGA eLIER 1-550578, as expected for a shock-dominated system \citep[e.g.,][]{Dopita2012}. However, we find that none of the LEGGOS sources follow a similar distribution, corroborated by a two-sided Kolmogorov-Smirnov test comparing the $v_{\text{shock}}/\text{FWHM}_{\ha}$ distributions for 1-550578 to each of the LEGGOS sources.  


We find the lack of evidence for shocks in the LEGGOS sources puzzling given the expectation for elevated shock signatures at $z\gtrsim2$, which we explore here. Given that the galaxies in our sample do not have dominant signatures of AGN photoionization (C. Luettgenau et al. in preparation will discuss the potential for sub-dominant accreting black holes in LEGGOS), the presence of any shock-excited gas would likely be due to processes related to stellar evolution. Shocks in predominantly star forming galaxies can have a substantial contribution to observed emission line fluxes \citep[e.g.,][]{Rich2014,Krabbe2014,Jaskot2016}. We consider a model of an ISM energized by shocks of non-AGN origin. For shocks driven entirely by stellar processes, the kinetic energy deposition into the ISM must be equal to the energy deposition from star formation, thus the energy injected into the ISM scales with the star formation rate \citep[e.g.,][]{Dekel2003,Chisholm2015}. Specific star formation rates are elevated at $z\sim2-4$ compared to low-redshift \citep[e.g.,][]{Cole2025}, including the LEGGOS galaxies and their clumpy \ha-emitting regions in particular. The star formation rate evolves closely with the core-collapse supernova rate, which peaks at the epochs traced by LEGGOS \citep[for recent discussion, see, e.g.,][]{DeCoursey2026,Vassallo2026}. Additionally, an increasingly top-heavy initial mass function, a scenario often leveraged at even higher redshifts \citep[e.g.,][]{Cameron2024}, would add to the rate of stellar shocks interacting with the ISM. Future work with a growing sample size and the development of new models will indicate if the LEGGOS galaxies are representative of shock properties in all galaxies at these redshifts. 

Major mergers are a means of shock production beyond stellar evolution and AGN \citep[e.g.,][]{Marketvitch2002,Marketvitch2007,Ha2018}. The one galaxy in LEGGOS which is identified as an active merger, SGAS 2111, shows evidence for evolved stellar populations with minimal nebular features in its spectrum (a full analysis of SGAS 2111 is ongoing in R. Oh et al. in prep.). This particularly includes no pixel-scale detections of the strongest shock indicator shown in Figures \ref{fig:ratios_leggos_part1} and \ref{fig:ratios_leggos_part2}, $\oi~\lambda6302$. While we find no evidence for merger induced shocks in SGAS 2111, we emphasize that any statements about the population level presence of merger induced shocks at and beyond cosmic noon will require larger samples of galaxies with IFU spectroscopy.

It is important to note that the regions of the emission line ratio diagnostics which select shocks can also be populated by diffuse ionized gas (DIG) \citep[e.g.,][]{Zhang2017}. Particularly, DIG can produce elevated low-ionization line ratios, e.g., \nii/\ha, \sii\ha, and \oi/\ha, but are likely only found in low surface brightness \citep[$\Sigma(\ha) < 10^{39}$ erg s$^{-1}$ kpc$^{-1}$;][]{Zhang2017} pixels outside of HII regions. This makes DIG inherently difficult to detect, particularly in high-redshift galaxies. Additionally, the ionization mechanisms driving DIG may be diverse, including Lyman continuum leakage from H II regions and AGN, to hot evolved low-mass stars, to radiative shocks \citep[e.g.,][]{Belfiore2022,Lagos2026}. While we do not specifically test for the presence of DIG in this work, we note that such diffuse gas could potentially contribute to some of the elevated \sii/\ha\ and \oi/\ha\ ratios observed in some spaxels in the LEGGOS sources. We leave a dedicated evaluation of the presence of DIG in the LEGGOS galaxies for future work (T. Hutchison et al. in prep.).

\subsection{Limitations and Improvements for Future Work}\label{sec:discussion:improvements}
While the shock model database from 3MdB$^S$ is relatively large ($\sim$200k total shock models), the grid is often sparse in some parameters (e.g., pre-shock density). Any choice of grid priors and model density directly impacts the parameter inference methods performed in Section \ref{sec:results} \citep[for similar discussion, see, e.g.,][]{Lebouteiller2022,Li2025,Cleri2025,Cleri2026}. The grids are sparsely modeled for pre-shock density, which impacts the reliability of our analysis of the inferred compression factor. Additionally, the choice of abundance patterns \citep[e.g.,][]{Allen.2008,Gutkin2016,Alarie2019}, is likely to become increasingly more impactful to the inference of ISM parameters at higher redshifts \citep[e.g.,][]{Alarie2019,Li2024,Cleri2025,Flury2025}.

Future work will benefit from spatially-resolved multiwavelength spectroscopy, particularly in the UV and IR, to differentiate between ionizing sources when shocks are suspected. Ultraviolet diagnostics with lines such as $\ciii~\lambda\lambda1907,1909$, $\heii~\lambda1640$, and $\oiiisemi~\lambda\lambda 1660,1666$ have shown some promise at identifying shocks through models and with integrated spectra at high redshift \citep[e.g.,][]{,Allen1998,Jaskot2016,Hirschmann2019,Calabro2022,Mingozzi2024,Flury2025}. While the LEGGOS sources do have archival ground-based rest-UV spectra \citep{Bayliss14,Johnson2017_1110paperI,Rigby_2018a,Khullar2026}, all the relevant diagnostic lines are only detected in two targets: SGAS1050 \citep{Bayliss14} and the Sunburst Arc \citep[e.g.,][]{mainali2022}. Additionally, the lack of spatial resolution in these long-slit spectra make the pixel-by-pixel shock search undertaken here practically impossible. Future rest-UV IFS observations of these or other lensed galaxies could help to improve future shock searches at Cosmic Noon. 

Near-IR line ratios leveraging the $\feii~1.257 \mu \text{m}$ line have also been noted as strong tracers of potential shocks at high redshifts \citep[e.g.,][]{Brinchmann2023,Calabro2023,Shapley2025}. The NIRSpec/prism observations from LEGGOS cover the \feii~$\lambda 1.257\mu\text{m}$ line for four of our targets, the Cosmic Eye, SGAS1429, SGAS1527, and SGAS1110, and we detect the line in the galaxy-integrated spectra of all but SGAS1110 (B. Welch et al., in prep). Examining the $\log($\feii/Pa-$\gamma)$ ratio in the galaxy-integrated spectra, we find ratios of $-0.93 \pm 0.02$, $-0.91 \pm 0.08$, and $-0.80\pm 0.02$ for SGAS1429, SGAS1527, and the Cosmic Eye, respectively. The elevated \feii/\pab\ observed in the Cosmic Eye is likely driven by metallicity, as that is the only galaxy in the LEGGOS sample with a super-solar oxygen abundance (B. Welch et al., in prep). These ratios appear fully consistent with photoionization from recent star formation, though we note that an explicit demarcation between shocks and other photoionization sources has yet to be published.

\section{Summary and Conclusions}\label{sec:summary}
In this work we present the first systematic search for shocks in $z\sim2-4$ galaxies using JWST/NIRSpec spatially-resolved spectroscopy for six galaxies from the LEGGOS survey. The primary findings of our work are as follows.

\begin{itemize}
    \item We find that none of the LEGGOS galaxies exhibit more than 8\% of spaxels in the image plane with emission line ratios consistent with shocks from the \cite{Kewley2006} diagnostics (Figures \ref{fig:ratios_leggos_part1} and \ref{fig:ratios_leggos_part2}). Of the shock-identified pixels, none exhibit spatial correlation in the image plane, indicating that they may be driven by scatter. The highest shock fractions are observed in \oi/\ha\ for SGAS1527 (7.9\%) and SGAS1110 (6.7\%), but these are based on the smallest number of measured spaxels. Of galaxy/diagnostic combinations with at least 100 measured spaxels, no galaxy exceeds a 3.3\% shock fraction, indicating that the LEGGOS galaxies are not significantly powered by shocks.
    \item We compare the velocity dispersions derived from \ha\ to the emission line ratios \nii/\ha, \sii/\ha, and \oi/\ha\ for narrow and broad components where measurable (Figures \ref{fig:velocity_dispersion_line_ratios_part1} and \ref{fig:velocity_dispersion_line_ratios_part2}). We find no evidence for a correlation between the velocity dispersions and the line ratios, further indicating that the LEGGOS galaxies do not contain significant fractions of shocked gas.
    \item We use MAPPINGS-V shock only and shock plus precursor models to infer shock velocities from the observed emission lines (Figure \ref{fig:vshock_fwhm_hist_six_panel}). We compare to MaNGA spectroscopy the shock-dominated ``extended LIER'' 1-550578 as low-redshift anchor. We find that the shock velocities inferred for the LEGGOS galaxies do not correlate with the velocity dispersions from the observed \ha\ as would be expected for shocked gas. The lack of these correlations further disfavor a shock interpretation for the LEGGOS sources.
\end{itemize}

This paper demonstrates the power of spatially-resolved spectroscopy of the rest-frame optical emission lines to diagnose the presence of shocks in lensed galaxies at $z\sim2-4$. The LEGGOS spectroscopy allow us to probe potential shocks at pixel, clump, and galaxy integrated scales for the first times at these epochs. While we do not find evidence for shocks in the LEGGOS galaxies, future IFU studies with larger samples are required to contextualize our sources in the broader galaxy population. 

\begin{acknowledgements}
    This work was supported by the International Space Science Institute (ISSI) in Bern, through the ISSI Visiting Scientist Program. The authors thank the construction crew digging a giant hole outside the ISSI offices for providing a consistently jarring soundtrack which kept us awake and alert while writing this paper. 

    This work has made use of and NASA’s Astrophysics Data System (ADS) Bibliographic Services and the Science Explorer (SciX).

    The authors thank the developers of the Mexican Million Models (3MdB) database for making their photoionization and shock models publicly available.
    
    This research made use of Marvin, a core Python package and web framework for MaNGA data, developed by Brian Cherinka, José Sánchez-Gallego, Brett Andrews, and Joel Brownstein. (MaNGA Collaboration, 2018).

    NJC acknowledges support from JWST-AR-05558, as well as funding from the Eberly Postdoctoral Fellowship from the Eberly College of Science at the Pennsylvania State University. NJC also thanks Jakob Helton for insightful discussion.

    TAH acknowledges support from an appointment to the NASA Postdoctoral Program at the NASA Goddard Space, administered by Oak Ridge Associated Universities under contract with NASA, as well as the University of Maryland Baltimore County and the Center for Space Sciences and Technology.

    The Pennsylvania State University campuses are located on the original homelands of the Erie, Haudenosaunee (Seneca, Cayuga, Onondaga, Oneida, Mohawk, and Tuscarora), Lenape (Delaware Nation, Delaware Tribe, Stockbridge-Munsee), Monongahela, Shawnee (Absentee, Eastern, and Oklahoma), Susquehannock, and Wahzhazhe (Osage) Nations. As a land grant institution, we acknowledge and honor the traditional caretakers of these lands and strive to understand and model their responsible stewardship. We also acknowledge the longer history of these lands and our place in that history.

\end{acknowledgements} 

\begin{contribution}
The first three authors contributed equally to this work. NJC, TAH, and BW: Conceptualization, data curation, formal analysis, investigation, methodology, visualization, writing - original draft, writing - review and editing. GK, MF, and MBB: Project administration. All authors: writing - review and editing.

\end{contribution}

\facilities{JWST/NIRSpec \citep{Gardner2006,Gardner2023,Jakobsen2022,boker2022}}

\software{
Astropy \citep{Astropy2013,Astropy2018}, 
CMasher \citep{vanderVelden2020},
IPython \citep{Perez2007},
MAPPINGS \citep{Dopita.2017,Sutherland.2017,Sutherland.2018}
Matplotlib \citep{Hunter2007}, 
NumPy \citep{Harris2020}, 
pandas \citep{Reback2022},
PyNeb \citep{Luridiana2015},
Seaborn \citep{Waskom2021},
SciPy \citep{Virtanen2020}
}

\bibliography{main}{}
\bibliographystyle{aasjournalv7}

\end{document}